\documentclass{ws-procs9x6}

\usepackage{psfrag}

\input{axodraw.sty} 
\newcommand{\beqn}   {\begin{eqnarray}}
\newcommand{\eeqn}   {\nonumber \end{eqnarray}}

\newcommand{\non}   {\nonumber \\[.1cm]}
\newcommand{\Dslash}{\mbox{$D$ \kern-.92em  \big /}}

\begin{document}

\title{
How Stueckelberg Extends \\ the Standard Model and the MSSM 
}

\author{
Boris K\"ors
}

\address{
Center for Theoretical Physics, 
Laboratory for Nuclear Science, and \\ 
Department of Physics,
Massachusetts Institute of Technology, \\ 
Cambridge, Massachusetts 02139, USA  \\ 
kors@lns.mit.edu
}

\author{Pran Nath}

\address{
Department of Physics, 
Northeastern University, \\ 
Boston, MA 02115, USA \\ 
nath@neu.edu
}  

\maketitle

\abstracts{
Abelian vector bosons can get massive through the Stueckelberg mechanism without 
spontaneous symmetry breaking via condensation of Higgs scalar fields. This appears very 
naturally in models derived from string theory and supergravity. The simplest 
scenarios of this type consist of extensions of the Standard Model (SM) or the
minimal supersymmetric standard model (MSSM) by 
an extra $U(1)_X$ gauge group with Stueckelberg type couplings. For the SM, the physical 
spectrum is extended by a massive neutral gauge boson Z$'$ only, while the extension of 
the MSSM contains a CP-even neutral scalar and two extra neutralinos. 
The new gauge boson Z$'$ can be very light compared to other models with $U(1)'$ extensions. 
Among the new features of the Stueckelberg extension of the MSSM, the most striking  
is the possibility of a new lightest supersymmetric particle (LSP)
$\tilde\chi_{\rm St}^0$ which is mostly composed of Stueckelberg fermions. 
In this scenario the LSP of the 
MSSM $\tilde\chi_1^0$ is unstable and decays into $\tilde\chi_{\rm St}^0$. Such decays alter 
the signatures of supersymmetry and have impact on  
searches for supersymmetry in accelerator experiments. Further, with R-parity 
invariance, $\tilde\chi_{\rm St}^0$ is the new candidate for dark matter. 
}

\section{Stueckelberg mechanism for gauge boson masses} 

The Stueckelberg mechanism\cite{stueck}, as an alternative to spontaneous symmetry breaking 
in the Higgs effect\cite{higgs}, describes a way to make the naive 
Lagrangian of a massive abelian vector boson $A_\mu$, such as 
\beqn
{ L}_{\rm St} = -\frac14 F_{\mu\nu}F^{\mu\nu} - \frac{m^2}{2} A_\mu A^\mu\ , 
\eeqn
gauge invariant. To achieve this, one replaces 
$A_\mu \longrightarrow A_\mu + \frac{1}{m} \partial_\mu \sigma$, where 
$\sigma$ is an axionic scalar that takes the role of the longitudinal mode of the massive 
vector, and defines the gauge transformation
\beqn
\delta A_\mu = \partial_\mu \epsilon \ , \quad \delta\sigma = -m\epsilon\ . 
\eeqn
The resulting Lagrangian is gauge invariant and renormalizable\cite{nonabelian,AML}. 
The physical spectrum  contains just 
the massive vector field, and this mass growth occurs without the need for a charged
scalar field developing a vacuum expectation value, without spontaneous symmetry breaking 
and  accordingly without the need for a Higgs potential. 
In the above the mass parameter $m$ is 
called a topological mass\cite{AML}. One can now go through the procedure of gauge fixing, 
so that $A_\mu$ and $\sigma$ decouple in the final theory, 
\beqn
{ L}_{\rm St} + { L}_{\rm gf} &=& -\frac14 F_{\mu\nu}F^{\mu\nu} - \frac{m^2}{2} A_\mu A^\mu
 - \frac{1}{2\xi} ( \partial_\mu A^\mu )^2 
-\frac12 \partial_\mu \sigma \partial^\mu \sigma
 - \xi \frac{m^2}{2} \sigma^2\ . 
\eeqn
A number of properties of this Lagrangian should be stressed. $i)$ The vector $A_\mu$ 
has absorbed the real scalar $\sigma$ in the process of getting a mass, with nothing left.
$ii)$ As the global subgroup of the gauge transformation, one can shift the scalar 
by a constant, $\delta \sigma = c$. This is a Peccei-Quinn like shift symmetry, and is the reason why 
we call $\sigma$ an axionic pseudoscalar, which only appears with derivative couplings.\footnote{But $\sigma$ 
does not necessarily have to couple to the QCD gauge fields in the usual topological term. In fact, we 
assume such couplings to be absent.}
$iii)$ Currently
it appears  possible to write such a gauge invariant Stueckelberg Lagrangian only for an 
abelian gauge symmetry, 
not for non-abelian gauge transformations\cite{nonabelian}. 
However, as will become clearer from the string theoretic embedding of the Stueckelberg mechanism 
into D-brane models, the relevant $U(1)$ gauge group can become a subgroup of 
some non-abelian and simple grand unified gauge group in higher dimensions. 

\section{Stueckelberg couplings in string theory and supergravity} 

One immediate way to see that Stueckelberg couplings appear in dimensional reduction of supergravity 
from higher dimensions, and in particular string theory, is to consider the reduction of the ten-dimensional 
$N=1$ supergravity coupled to supersymmetric Yang-Mills gauge fields, in the presence of 
internal gauge fluxes. The ten-dimensional kinetic term for the anti-symmetric 2-tensor $B_{IJ}$ 
involves a coupling to the Yang-Mills Chern-Simons form, schematically 
$\partial_{[I} B_{JK]} + A_{[I} F_{JK]} +\, \cdots$, in proper units. 
Dimensional reduction with a vacuum expectation value for the internal gauge field strength, 
$\langle F_{ij} \rangle \not=0$, leads to 
\beqn
\partial_{\mu} B_{ij} + A_{\mu} F_{ij} +\, \cdots 
~\sim~ \partial_\mu \sigma + m A_\mu  \ , 
\eeqn
after identifying the internal components $B_{ij}$ with the scalar $\sigma$ and the value of the 
gauge field strength with the mass parameter $m$, which is indeed a topologiucal quantity, related to the 
Chern numbers of the gauge bundle. 
Thus $A_\mu$ and $\sigma$ have a Stueckelberg coupling of the form
$A_\mu \partial^\mu \sigma$.  
These couplings play an important role in the Green-Schwarz anomaly cancelation mechanism. 
In the effective four-dimenional theory, for instance abelian factors in the gauge group can have 
an anomalous matter spectrum, whose ABJ anomaly is canceled by Green-Schwarz type contributions. These involve the two 
terms 
$m A^\mu \partial_\mu \sigma + c\, \sigma 
F_{\mu\nu} \tilde F^{\mu\nu}$
in the Lagrangian. 

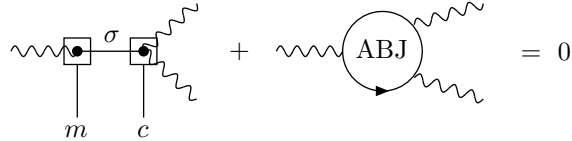
\begin{figure}[ht]
\hspace{.3cm}
\begin{picture}(300,70)

\Photon(50,50)(75,50){2}{4}
\Photon(100,50)(120,68){2}{4}
\Photon(100,50)(120,32){2}{4}
\Line(75,50)(100,50)
\Vertex(75,50)2
\Vertex(100,50)2
\EBox(70,45)(80,55)
\EBox(95,45)(105,55)
\Line(75,25)(75,45)
\Line(100,25)(100,45)

\Photon(150,50)(175,50){2}{4}
\ArrowArc(190,50)(15,90,450)
\Photon(202,60)(228,68){2}{4}
\Photon(202,40)(228,32){2}{4}

\Text(220,50)[l]{\hspace{-3.3cm} $+$}
\Text(410,50)[l]{\hspace{-7.2cm} $~=~0$}

\Text(21,20)[1]{$m$}
\Text(21,20)[1]{$c$} 
\Text(-16,56)[1]{$\sigma$}

\Text(61,51)[1]{ABJ}

\end{picture}
\end{figure}

\vspace{-.8cm}
As can be read from the left Feynman-diagram, the contribution to the anomalous 3-point function is 
proportional to the product of the two couplings, $m \cdot c$, while the mass parameter in the 
Stueckelberg coupling is only $m$.
Therefore, any anomalous $U(1)$ will always get massive through the Stueckelberg mechanism, 
since $m \cdot c\not=0$, but a non-anomalous $U(1)$ can do so as well, if $m \not=0,\, c=0.$ 
Since we do not want to deal with anomalous gauge symmetries here, we shall always assume that 
$m \not=0,\, c=0.$ The mass scale that determines $m$ 
within models that derive from string theory can, at leading order, 
also be derived from dimensional reduction. It turns out to be proportional to the string or 
compactification scale in many cases\cite{ghiletal}, 
but can in principle also be independent\cite{Ibanez:1998qp}. \\ 

The fact that an abelian gauge symmetry, anomalous or non-anomalous, may decouple from the 
low energy theory via Stueckelberg couplings was actually of great importance in the construction 
of D-brane models with gauge group and spectrum close to that of the 
SM\cite{Ibanez:2001nd}. 
Roughly speaking, these D-brane constructions start with a number of unitary gauge group factors, 
which are then usually broken to their special unitary subgroups via Stueckelberg couplings of all 
abelian factors, except the hypercharge, 
\beqn
U(3) \times U(2) \times U(1)^2
~~\stackrel{\rm Stueckelberg}{\longrightarrow}~~ SU(3)\times SU(2)_L \times U(1)_Y
\eeqn
The mass matrix for the abelian gauge bosons is then block-diagonal, and only the SM 
survives. 
In order to ensure this pattern, one has to impose an extra condition on the Stueckelberg mass 
parameters, beyond the usual constraints that follow from the RR charge cancellation constraints, 
namely that the hypercharge gauge boson does not couple to any axionic scalar 
and remains massless\cite{Ibanez:2001nd}. 
In the language of these D-brane models, we will here relax this extra condition, and allow the 
hypercharge gauge boson to have Stueckelberg type couplings, and thus mix with other abelian 
gauge factors beyond the SM gauge group, which seems a very natural extension of the 
SM in this frame work. 

\section{Minimal Stueckelberg extension of the Standard Model} 
To keep things as simple as possible, and study the essence of the Stueckelberg effect in its 
minimal version\cite{kn,Kors:2004ri}, 
we consider an extension of the SM with only one extra abelian gauge factor 
$U(1)_X$. Along with the $U(1)_X$ we also allow a hidden sector with charged matter fields. 
Since the Stueckelberg mechanism cannot break the non-abelian $SU(2)_L$ in the 
process of electro-weak symmetry breaking, we cannot replace the usual Higgs mechanism, but only 
add the Stueckelberg masses for the abelian gauge bosons of hypercharge and $U(1)_X$ on 
top of the usual Higgs mechanism. 
Schematically, the couplings are thus given by 
\beqn
&& SU(3)\times \overbrace{SU(2)_L\times U(1)_Y}^{{\rm Higgs}\ \Phi}\times U(1)_X \non
&& SU(3)\times SU(2)_L\times \underbrace{U(1)_Y\times U(1)_X}_{{\rm Stueckelberg}\ \sigma} \ . 
\eeqn
The extra degrees of freedom beyond the SM spectrum are then the new gauge boson $C_\mu$ plus the 
axionic scalar $\sigma$ which combine into the massive neutral gauge field Z$'$. 
Explicitily, we add the Lagrangian 
\beqn 
{ L}_{\rm St} &=& -\frac{1}{4} C_{\mu\nu}C^{\mu\nu} + g_X
C_\mu J^\mu_X - \frac{1}{2} (\partial_{\mu}\sigma  + M_1 C_{\mu} +
M_2 B_{\mu})^2
\eeqn
to the relevant part of the SM Lagrangian 
\beqn
{ L}_{\rm SM} &=& -\frac{1}{4} {\rm tr}\, F_{\mu\nu}F^{\mu\nu} -\frac{1}{4} B_{\mu\nu}B^{\mu\nu}
+ g_2 A^a_\mu J^{a\mu}_2 + g_Y B_\mu J^\mu_Y \non
&&
- D_{\mu}\Phi^{\dagger} D^{\mu}\Phi - V(\Phi^{\dagger}\Phi) + \
\cdots\ . 
\eeqn 
The scalars and vectors decouple after adding gauge fixing terms in a standard fashion. 
To keep the model as simple as possible, we impose the following constraints on the charged matter 
spectrum: We assume that the fermions of the SM are neutral under $U(1)_X$, and  vice versa 
we require that all the  
fields of the hidden sector, which are potentially charged under $U(1)_X$, be neutral under the 
SM gauge group.\footnote{These conditions are very hard to satisfy in computable D-brane models, such 
as intersecting D-brane models with tori or toroidal orbifolds as compactification space. 
As a matter of principle, this should be an artifact of the too simple toroidal geometry and 
topology, and we expect that one can get around this problem in more general Calabi-Yau 
compactifications.}
Finally, one has to make sure that the hidden sector dynamics can really be ignored, 
i.e.\ that there is no spontaneous breaking of the $U(1)_X$ in the hidden sector.

\section{Stueckelberg effects in the Standard Model} 

All effects of the minimal Stueckelberg extension 
on the SM Lagrangian can be summarized by the modified mass matrix of the, 
now three, neutral gauge bosons. In the basis $(C_{\mu}, B_{\mu}, A_{\mu}^3)$ for $U(1)_X$, hypercharge 
and the 3-component of iso-spin, The vector boson mass matrix is  
\beqn
\left[
\begin{array}{c|cc}
M_1^2  &  M_1M_2  &  0\\
\hline
M_1M_2 & M_2^2 + \frac{1}{4} g_Y^2 v^2 & - \frac{1}{4}g_Yg_2 v^2 \\
0 & -\frac{1}{4}g_Yg_2 v^2 & \frac{1}{4}g_2^2 v^2
\end{array}
\right]\ . 
\eeqn
The mass matrix above has one massless eigenstate, the photon with $M^2_\gamma =0$, and two 
massive ones with eigenvalues
\beqn
M^2_{\rm Z} = \frac{v^2}{4} ( g_2^2 + g_Y^2 ) + \, {\rm O}(\delta)\ , 
\quad M^2_{\rm Z'} = M^2 + {\rm O}(\delta)\ . 
\eeqn
 In the above we have introduced two parameters $M$ and $\delta$, defined by
$M^2=M_1^2+M_2^2$ and $\delta =M_2/M_1$ 
 to parametrize the Stueckelberg extension, 
which are
a mass scale and a small coupling parameter. The bounds for these two, the values for which the Stueckelberg 
extension is still safely within experimentally allowed ranges, have been given in\cite{kn} as 
\beqn
M > [150\, {\rm GeV}] \ , \quad \delta  < 0.01 \ . 
\eeqn
For later use, we also define $M_0^2 = v^2( g_2^2 + g_Y^2 )/4$. 
Roughly speaking, one may notice that all couplings which allow communication of SM fields to the 
Stueckelberg sector, are suppressed by $\delta$. In models where an extra $U(1)'$ addition to the 
SM gauge group is broken by a Higgs condensate, the suppression factor comes from the propagator of the massive 
gauge boson, and is of the order of $M_{\rm Z}^2/M_{{\rm Z}'}^2 \sim 0.01$\cite{cl2,barger,rizzo}. 
This demands that the mass of 
the new gauge boson has to be in the range of TeV, much larger than the bound on $M$ as above. For the 
Stueckelberg model, the 
suppression can easily be achieved by demanding $\delta$ be 
sufficiently small.\footnote{To date a real global fit of the experimental data 
within the Stueckelberg extended model has not been performed, and thus these 
statements cannot be made completely quantitative as yet.} \\ 

A regime in the parameter space which would also be very interesting to investigate is the region 
%
$M_1,\, M_2 \rightarrow \infty ,\ \delta = {\rm finite}.$
%
This scaling limit corresponds to the expected behaviour in string theoretic models with a high mass scale, but 
basically arbitrary coupling $\delta$. One may speculate that even though the Z$'$ becomes very heavy, and 
cannot be produced directly, small effects may remain observable. In essence, the Z$'$ does not decouple from 
the low energy theory, because the mass matrix is not diagonal, and this may have important consequences. \\ 

To proceed further, one can now diagonalize the vector boson mass matrix by 
an orthogonal matrix $O = O (\theta,\phi,\psi)$,  
a function of three angle variables $\{\theta, \phi, \psi\}$, 
%
%
with (see\cite{kn} for the details of the notation) 
\beqn
\tan(\phi) = \delta\ , \quad
\tan (\theta) ~=~ \frac{g_Y}{g_2}\cos(\phi) \ , \quad
\tan (\psi) = \frac{g_2(1-M^2/M_{{\rm Z}'}^2)}{g_Y \sin(\phi) \cos(\theta)}\ . 
\eeqn
The bounds on $M$ and $\delta$ translate into $\phi, \psi < 1^0$ and $\theta \sim \theta_W$, 
which is the electro-weak mixing angle. 
Inserting the mass eigenstates into the interaction Lagrangian 
\beqn
{ L}_{\rm int} = g_2 A_\mu^a J^{a\mu}_2 + g_Y B_\mu J^\mu_Y +
g_X C_\mu J^\mu_X
\eeqn
one finds the interactions of the phyical vector fields. As the most striking example one 
gets for the photonic interactions  
\beqn
e A_\mu^\gamma J^\mu_{\rm em} =
\frac{g_2g_Y\cos(\phi)}{\sqrt{g_2^2+g_Y^2
\cos^2(\phi)}} A_\mu^\gamma
  \Big( J_Y^\mu + J_2^{3\mu} - \frac{g_X}{g_Y} \tan(\phi) J^\mu_X \Big)\ , 
\eeqn
which implies that the electric charge unit is now slightly redefined by 
\beqn
e = \frac{g_2g_Y\cos(\phi)}{\sqrt{g_2^2+g_Y^2 \cos^2(\phi)}}
\eeqn
and that the charge unit of fields in the hidden sector, with charge under $U(1)_X$, is 
irrational and very small.\footnote{Note that a hidden sector with such an irrational 
electric charge is a possible consequence of the Stueckelberg scenario, which survives as a 
potentially observable fact in the stringy limit $M \rightarrow \infty,$ $\delta ={\rm finite}$. 
While the mass of the gauge boson Z$'$ becomes very large, the hidden sector charged matter 
fields can be massless at the string scale, and thus be much lighter and in the end observable 
in experiment.} 
Since all exotic couplings are suppressed by powers of $\delta$, computable finite quantities 
that lead to definite predictions are best found in terms of ratios of such couplings. 
Examples are the branching ratios of the Z$'$, or the forward-backward asymmetry at the 
Z$'$ peak in resonant producation via $e^+ e^-$ collision\cite{kn}. 
One finds for example that the total width of the Z$'$ is extremely small,
at least  as long as the hidden sector 
charged fields are heavy enough, i.e.,
\beqn
\Gamma({\rm Z}'\rightarrow f\bar f) ~\sim~ {\rm O}(10)\, {\rm MeV}\ . 
\eeqn
Therefore, Z$'$ would appear as a very sharp peak in $e^+e^-$ annihilation
and in other collider data.

\section{Stueckelberg extension of the MSSM: StMSSM} 

The supersymmetrized version of the Stueckelberg coupling is related to the so-called linear 
multiplet formalism, see\cite{kleinetal}. For the present minimal 
Stueckelberg extension of the MSSM, which we call StMSSM, it reads\cite{Kors:2004ri}
\beqn
{ L}_{\rm St} = \int d^2\theta d^2\bar \theta\ (M_1C+M_2B+  S +\bar S )^2 \ . 
\label{mass}
\eeqn
Here $S$ is the Stueckelberg chiral multiplet, and $B,\ C$ are the abelian vector
multiplets 
of hypercharge and $U(1)_X$. 
The degrees of freedom in components are given by  
%
$S = (\chi,\rho+i\sigma,F), \ B=(B_\mu,\lambda_B,D_B)$,
and $\ C=(C_\mu,\lambda_C,D_C),$
%
and the Lagrangian becomes 
\beqn
{ L}_{\rm St} &=& - \frac{1}{2}(M_1C_{\mu} +M_2 B_{\mu} +\partial_{\mu} \sigma)^2
 - \frac{1}{2} (\partial_\mu \rho)^2
 -\frac{i}{2} (\chi \sigma^{\mu}
 \partial_{\mu}\bar \chi - (\partial_{\mu}\chi)\sigma^{\mu} \bar\chi) 
\non
&&  +\rho(M_1D_C +M_2 D_B)
 +[ \bar \chi (M_1\bar \lambda_C + M_2\bar \lambda_B)
 + {\rm h.c.} ] +2|F|^2 \ . 
\eeqn
One may of course also add Fayet-Iliopoulos terms for each one of the two abelian factors, 
which would contribute to the scalar potential. As is usually done for the MSSM, we will however 
assume that their contributions to the breaking of supersymmetry are small 
compared to other sources, and therefore neglect these terms throughout, leaving a more 
complete analysis to future work. 
Eliminating the auxiliary fields $F, D_B, D_C$ one thus finds corrections to the usual D-term potential 
of the MSSM through the coupling of the D-fields to $\rho$.
To complete the action, we add the following soft supersymmetry breaking terms
for the neutral gauginos, and the scalars $\{h_1, h_2, \rho\}$, 
\beqn
{ L}_{\rm soft} &=& -\frac12 \tilde m_\rho^2 \rho^2 -\frac12
\tilde m_1 \bar \lambda_B \lambda_B -\frac12 \tilde m_X \bar
\lambda_C \lambda_C
\non
&&
-\frac12 m_1^2 |h_1|^2
-\frac12 m_2^2 |h_2|^2 - m_3^2 ( h_1 \cdot h_2 + \ {\rm h.c.}\ )\ . 
\eeqn
An important difference between the Stueckelberg Lagrangian and the Lagrangian that involves
the Higgs mechanism is 
the fact that the chiral fermion $\chi$ is neutral under the gauge group. A scalar Higgs 
condensate would need to be charged under $U(1)_X$ to break the gauge symmetry spontaneously. But then, 
the standard coupling of the fermionic partner $\tilde h$ of the Higgs scalar in the form 
$g_Y B_\mu \tilde h\sigma^\mu \bar{\tilde h}, \
g_X C_\mu \tilde h\sigma^\mu \bar{\tilde h}$ would imply a contribution ot the ABJ triangle anomaly, 
and a second Higgs multiplet of opposite charges would be needed to cancel the anomaly, just as in the 
MSSM. This is not the case for the Stueckelberg mechanism, where $\chi$ does not have such couplings, 
and no second chiral multiplet is needed. \\ 

Putting things together, the new effects in the StMSSM are 
in the scalar potential and the neutralino mass matrix, in addition to the mass matrix of 
the neutral gauge bosons, which is identical to that of the extended SM. 
The scalar potential with the three CP-even scalars $\rho,\ h_1,\, h_2$ is 
\beqn
{V}(h_1,h_2,\rho) &=& \frac{1}{2} (M_1^2+M_2^2+\tilde m_{\rho}^2)
\rho^2 + V_D^{\rm MSSM} (h_1,h_2)
\non
&& \hspace{-2cm}
+ \frac{1}{2} ( m_1^2 - \rho g_Y M_2 ) |h_1|^2 + \frac{1}{2}
(m_2^2 + \rho g_Y M_2 ) |h_2|^2 + m_3^2 (h_1 \cdot h_2 +\ {\rm
h.c.}\ )\ , 
\eeqn
where $V_D^{\rm MSSM} (h_1,h_2)$ is the usual D-term potential of the MSSM for the
Higgs fields $h_1, h_2$ of MSSM.
The new scalar $\rho$ modifies the Higgs mass terms through its vacuum expectation value. 
Shifting $\rho\rightarrow  v_\rho + \rho$ with $|g_Y M_2 v_\rho| < 10^{-4}\, M_{\rm Z}^2$, 
one however finds that this induces only very tiny effects, as for instance in the 
electro-weak symmetry breaking constraint 
\beqn
\frac12 M_{0}^2
=\frac{m_1^2-m_2^2\tan^2(\beta)}{\tan^2(\beta) -1} + \frac{g_Y M_2 v_\rho}{\cos (2\beta)} \ . 
\eeqn
The CP-even scalar mass matrix in the basis $(h_1, h_2, \rho)$ reads 
\beqn
\left[
\begin{array}{cc|c}
M_{0}^2c^2_{\beta} +m_A^2 s^2_{\beta}
 &  -(M_{0}^2+m_A^2)s_{\beta}c_{\beta}  & -t_{\theta}c_{\beta} M_{\rm W} M_2 \\
 -(M_{0}^2+m_A^2)s_{\beta}c_{\beta}& M_{0}^2s^2_{\beta} +m_A^2 c^2_{\beta}&
t_{\theta}s_{\beta} M_{\rm W} M_2  \\
\hline -t_{\theta}c_{\beta} M_{\rm W} M_2  & t_{\theta}s_{\beta}
M_{\rm W} M_2 & M^2 + \tilde m_{\rho}^2
\end{array}
\right] \ .
\nonumber
\eeqn
Going through the details, one again finds a very narrow resonance for the third mass 
eigenstate $\rho_S$ in the $J=0^+$ channel, similar to the Z$'$, i.e., 
\beqn
\Gamma( \rho_S \rightarrow t\bar t\, ) ~\sim~ {\rm O(10)\, MeV}\ . 
\eeqn
Perhaps the most interesting sector of the StMSSM is the fermionic neutralino mass matrix, 
which now involves the usual two higgsinos and two gauginos of MSSM, plus the new gaugino 
of $U(1)_X$ and 
the chiral fermion $\chi$ of the Stueckelberg multiplet. In the basis 
$(\chi, \lambda_C, \lambda_B, \lambda_3, \tilde h_1, \tilde h_2)$ it reads
\beqn
\left[
\begin{array}{cc|cccc}
0 & M_1 & M_2 & 0 & 0 & 0\\
M_1& \tilde m_X & 0 & 0 & 0 & 0 \\
\hline
M_2& 0 & \tilde m_1 & 0 & -c_1M_0 & c_2M_0\\
0 & 0 & 0 & \tilde m_2 & c_3M_0 & -c_4M_0 \\
0 & 0 & -c_1M_0  &  c_3M_0 & 0 & -\mu \\
0 & 0 & c_2M_0  &  -c_4M_0 &  -\mu & 0
\end{array}
\right]
\eeqn
with abbreviations 
$c_1=c_{\beta}s_{\theta}$, $c_2=s_{\beta}s_{\theta}$,
$c_3=c_{\beta}c_{\theta}$, $c_4=s_{\beta}s_{\theta}$, further $s_\theta$, $s_\beta$, etc., 
standing for the sin and cos of $\theta_W$ and $\beta$, 
where $\tan(\beta) =\langle h_2\rangle /\langle h_1\rangle$. 
It is convenient to number the eigenstates of the $4\times 4$ MSSM mass matrix according to 
$m_{\tilde \chi_1^0}<m_{\tilde\chi_2^0}<m_{\tilde\chi_3^0}<m_{\tilde\chi_4^0}$, and 
the two new ones by
\beqn
m_{\tilde\chi_5^0},\ m_{\tilde\chi_6^0} ~=~ \sqrt{M_1^{2}
+\frac{1}{4}\tilde m_X^{2}} \pm\frac{1}{2} \tilde m_X +
{\rm O}(\delta) \ , \quad m_{\tilde\chi_5^0}\ge m_{\tilde\chi_6^0}
\nonumber\ .
\eeqn
The parameter space now easily allows the situation that the lighter one among the two new neutralinos 
becomes the lightest among all six. It therefore is the LSP, and with R-parity conservation, the  natural dark matter 
candidate of the model. In this case, when $m_{\tilde\chi^0_{6}}<m_{\tilde\chi_1^0}$, we call $\tilde\chi^0_{6}$ 
the Stueckelino $\tilde\chi_{\rm St}^0$, and important modifications of the usual signatures for searches 
for supersymmetry in accelerator experiments\cite{trilep} would follow. One would actually observe decay 
cascades (see also\cite{Ellwanger:1997jj}), 
in which the lightest neutralino of the MSSM would further decay into the true LSP, the Stueckelino, via 
emitting fermion pairs, 
\beqn
\tilde\chi_1^0~\rightarrow~ {l}_i\bar {l}_i \tilde\chi_{\rm St}^0\ , \quad
q_j\bar q_j \tilde\chi_{\rm St}^0\ ,\quad
{\rm Z} \tilde\chi_{\rm St}^0\ . 
\eeqn
This would lead to decays of the form 
\beqn
\tilde {l}^{-} ~\rightarrow~ l^-+ \tilde\chi_{1}^0 ~\rightarrow~
{l^-} + \left\{ {  l_i^-l_i^+ + \{\tilde\chi_{\rm St}^0 \} \atop
                  q_j\bar q_j + \{\tilde\chi_{\rm St}^0 \} } \right.
\nonumber
\eeqn
for sleptons or 
\beqn
\tilde\chi_1^{-}~\rightarrow~ l^- + \tilde\chi_1^0 +\bar \nu
~\rightarrow~ l^- + \left\{ { l_i^-l_i^+ + \{\tilde\chi_{\rm St}^0 + \bar\nu \}
                    \atop q_j\bar q_j + \{\tilde\chi_{\rm St}^0 + \bar\nu \} } \right.
\eeqn
for charginos, and similarly for squarks and gluinos. 
The analysis above shows that there will be multilepton final states 
in collider experiments which would be a characteristic signature for this 
of scenario.

\section{Summary} 

We summarize now the most important features of the minimal Stueckelberg 
extension of the SM and of the MSSM: 
\begin{enumerate}
\item
The Stueckelberg mechanism provides a gauge invariant, renormalizable method 
to generate masses for abelian gauge bosons with minimal extra residual scalar 
fields in the system. The mass of the massive vector boson is ``topological''
in nature.
\item 
It naturally appears in many models that descend from string theory and higher-dimensional 
SUGRA. 
\item 
The Stueckelberg extension is very economical and distinct, even 
at the level of the degrees of freedom, compared to 
Higgs models with extra $U(1)'$ gauge factors.
\item 
In the Stueckelberg extension of SM, only the vector boson sector is
affected, as it introduces an extra Z$'$ boson which is typically a very narrow resonance.
 In addition, one has small exotic couplings of the photon and the
 Z with hidden matter, if present.
\item 
In the Stueckelberg extension of the MSSM, the vector boson sector, 
the Higgs sector and the neutralino sectors are affected. 
The effect on the vector boson sector is identical to what one has in
the Stueckelberg extension of the SM. In the neutral CP even Higgs sector 
one has mixing among the two CP-even Higgs of MSSM and a new CP-even
Stueckelberg scalar field $\rho$. In the neutralino sector, one finds
two more neutralinos which are mostly mixtures of neutral Stueckelberg
fermions, in addition to the four neutralinos of MSSM. 

\end{enumerate}

\section*{Acknowledgements} 

The work of B.~K.~was supported by the German Science Foundation (DFG) and in part by
funds provided by the U.S. Department of Energy (D.O.E.) under cooperative research agreement
$\#$DF-FC02-94ER40818.  The work of P.~N. was supported in part by
the U.S. National Science Foundation under the grant NSF-PHY-0139967.

\end{document}